\documentclass[twocolumn,notitlepage]{revtex4-1}
\usepackage{bm}
\usepackage{graphicx}
\usepackage{amsmath}
\usepackage{amssymb}
\usepackage{amscd}
\usepackage{epstopdf}
\usepackage{verbatim}
\usepackage{color}
\usepackage{array}
\usepackage{xspace}
\usepackage{xstring}
\usepackage{layouts}

\newcommand{\ie}{i.e.\xspace}
\newcommand{\eg}{e.g.\xspace}
\newcommand{\rhox}{\rho}
\newcommand{\rhof}{p}
\newcommand{\convex}[1]{\bar{#1}}
\newcommand{\rhofc}{\convex{p}}
\newcommand{\rhoxc}{\convex{\rho}}
\newcommand{\myskip}{\medskip}
\definecolor{greenA}{RGB}{0,190,0}

\begin{document}

\title{Self-propelled particle in a nonconvex external potential: \\ Persistent limit in one dimension}
\author{Yaouen Fily}
\affiliation{Wilkes Honors College, Florida Atlantic University, Jupiter FL 33458, USA}

\begin{abstract}
Equilibrium mapping techniques for nonaligning self-propelled particles have made it possible to predict the density profile of an active ideal gas in a wide variety of external potentials, however they fail when the self-propulsion is very persistent and the potential is nonconvex, which is precisely when the most uniquely active phenomena occur. Here we show how to predict the density profile of a 1D active Ornstein-Uhlenbeck particle in an arbitrary external potential in the persistent limit and discuss the consequences of the potential's nonconvexity on the structure of the solution, including the central role of the potential's inflection points and the nonlocal dependence of the density profile on the potential.
\end{abstract}

\maketitle


\section{Introduction}

Active matter is a class of driven nonequilibrium systems in which the driving forces are controlled locally rather than globally. The development of a unified theoretical framework for active matter has transformed the way we understand a wide range of biological systems (\eg, the cell's cytoskeleton, bacterial colonies, animal flocks) and allowed the emergence of new types of biomimetic materials (\eg, self-propelled colloids, vibrated grains, motor-filament suspensions)~\cite{Marchetti2013}. On the other hand, the nonequilibrium nature of active systems makes it extremely challenging to provide exact answers to even the most basic questions. 

Active systems often exhibit strong, counter-intuitive responses to confinement including rectification, curvature-dependent attraction towards walls, spontaneous motion and deformation of passive objects immersed in or containing an active fluid, lack of an equation of state for pressure, dynamic geometric frustration, and more~\cite{Galajda2007,DiLeonardo2010,Sokolov2010,Kaiser2014,Guidobaldi2014,Mallory2014,Fily2014a,Fily2015,Fily2016,Fily2017,Mallory2015,Nikola2016,Yan2015,Takatori2016,Solon2015,Fily2018,Keber2014,Sknepnek2015}.
Yet, the density profile of a single active particle in an external potential in unknown in the general case, even in one dimension. In some cases the problem can be mapped onto an equilibrium problem and exact or good approximate results exist~\cite{Tailleur2009,Nash2010,Maggi2015}. Still, some of active systems' most striking behaviors occur precisely in the regime where this technique fails~\cite{Fily2017}. In this paper, we show how to derive exact analytical results for the density profile in an external potential in the persistent regime where current equilibrium mapping techniques least apply. 

A number of recent advances on mapping active systems onto equilibrium ones have been driven by the Gaussian colored noise model of self-propulsion, also known as the active Ornstein-Uhlenbeck particle (AOUP) model. It allows the mapping and subsequent approximations to be done in a well controlled manner in a wide variety of systems~\cite{Farage2015,Maggi2015,Marconi2015,Wittmann2017,Wittmann2017a}. Its biggest caveat is that the derivation of steady-state quantities such as the density profile relies on the positive-definiteness of the operator $(1+\tau\mu \mathcal{H})$ where $\tau$ is the persistence time, \ie, the correlation time of the self-propulsion force, $\mu$ is the mobility, and $\mathcal{H}$ is the Hessian of the potential. If the violations of this condition are mild, or if they only occur in regions the particles hardly visit, it is possible to mend the theory and make predictions based on a modified equilibrium mapping~\cite{Fily2017,Wittmann2017}. Conversely, strongly negative eigenvalues of $(1+\tau\mu \mathcal{H})$ in regions the particles do explore can result in unusual properties, \eg, an ideal active gas whose density profile depends nonlocally on the external potential~\cite{Fily2017}. Interestingly, Ref.~\cite{Fily2017} derives the exact density profile in a situation in which $(1+\tau\mu \mathcal{H})$ has infinitely negative eigenvalues in locations that matter, about as far from the normal range of validity of the equilibrium mapping approach as it gets. It does so by restricting itself to the infinite persistence limit and using a quasistatic approximation wherein the self-propulsion force and the force exerted by the wall always cancel, effectively slaving the spatial dynamics to that of the self-propulsion force~\cite{Fily2014a,Fily2015,Fily2016}. 

In this paper, we show that a similar quasistatic approximation can be used to predict the steady-state density profile of a self-propelled particle in an arbitrary external potential in one dimension in the persistent limit. The predicted density profile in turn exhibits some of the same features previously predicted under nonconvex hard-wall confinement~\cite{Fily2015,Fily2017}. We first describe the AOUP model and the meaning of the persistent ($\tau\rightarrow\infty$) and quasi-thermal ($\tau\rightarrow0$) limits (section~\ref{model}). We then discuss the quasistatic approximation and its generalization from hard-wall confinement~\cite{Fily2015,Fily2017} to arbitrary external potentials (section~\ref{quasistatic}). In convex potentials, we show that the quasistatic approach predicts a density profile obtained by stripping the equilibrium mapping prediction of every term that survives in the quasi-thermal limit (section~\ref{convex}). In nonconvex potentials, we explicitly derive the density profile in a double well potential (equation~\eqref{eq:doublewell_prediction}), then describe an algorithm to derive the density profile in an arbitrary confining potential (section~\ref{nonconvex}). Finally in section~\ref{discussion} we discuss three key features of the density profile in nonconvex external potentials: the emptiness of concave regions, the crucial importance of the potential's inflection points, and the nonlocal dependence of the profile on the potential.


\section{Model}
\label{model}

We consider an active Ornstein-Uhlenbeck particles (AOUP) with position $x$ and self-propulsion force $f$ in an external potential $U$ in one dimension. The equations of motion are
\begin{align}
\dot{x} & = \mu [f - U'(x)] 
\label{eq:eom1} \\
\dot{f} & = -\frac{f}{\tau} + \frac{\sqrt{D}}{\mu\tau}\, \xi(t)
\label{eq:eom2}
\end{align}
Dots and primes denote time and space derivatives, respectively. The first equation describes an overdamped particle with mobility $\mu$ and self-propulsion force $f$ in an external potential $U$. The second equation describes an Ornstein Uhlenbeck process, \ie, a random walk in a harmonic potential. $\tau$ is the persistence time. It is the typical time it takes $f$ to change significantly. $D$ is the active diffusion constant. The long-time ($t\gg\tau$) behavior of a free particle ($U=0$) is diffusive with diffusion constant $D$. $\xi$ is a white Gaussian noise with zero mean and variance $\langle\xi(t)\xi(t')\rangle=2\delta(t-t')$ where $\delta$ is the Dirac delta function. 

Three additional quantities, derived from the ones above, are particularly useful to our analysis: the AOUP's root mean squared self-propulsion force $f_0=\sqrt{D/(\tau\mu^2)}$, 
its root mean squared velocity in the absence of external force $v_0=\mu f_0$, also called active velocity, and its persistence length $\ell=v_0\tau$.

For convenience we work in a unit system in which $\mu=1$ and omit $\mu$ in the rest of the paper. This effectively makes velocities and forces interchangeable. 
\myskip
 
\emph{Single active particle vs. ideal active gas.} 
The statistical properties of a collection of noninteracting active particles, or ideal active gas, follow straightforwardly from those of a single particle. Typically, the dynamics is better understood in terms of a single particle whereas the statistics, including the density profile, are better understood in terms of a collection of particles. In this paper we use the two points of view interchangeably, choosing whichever we believe makes a specific point easier to understand. It should be kept in mind that, when referring to multiple particles, we always mean noninteracting particles. Furthermore, all normalization factors are meant for a single particle. 
\myskip

\emph{Thermal limit, persistent limit, and relevant variables.} 
Only two of $D$, $\tau$, $f_0$ are needed to characterize the active noise. Which two are most relevant depends on the regime one is interested in. 

When the persistence time $\tau$ is negligible, the active noise $f(t)$ is equivalent to a thermal noise with temperature $T=D/(\mu k)$ where $k$ is the Boltzmann constant.
In this case $D$ is the only relevant variable and $f_0=\sqrt{D/\tau}$ is infinite. If $\tau$ is small but not negligible, $D$ and $\tau$ are the right variables for a perturbative expansion: $D$ defines the thermal solution to expand about while $\tau$ is the expansion parameter.

The main purpose of this paper is to derive the density profile in the opposite limit, or persistent limit, when $\tau$ is large compared to every other time scale in the system. We choose to derive this density profile in terms of $D$ and $\tau$ in order to 1) remain consistent with equations~\eqref{eq:eom1}-\eqref{eq:eom2}, which are our starting point, and 2) simplify the comparison with the thermal limit. However, it should be noted that the final result (equations~\eqref{eq:profile_convex} and \eqref{eq:summary1}-\eqref{eq:summary3}) only depends on $D/\tau$, \ie, on $f_0$. In particular, the persistent limit should be taken at constant $f_0$, with both $\tau$ and $D=\tau f_0^2$ going to infinity.


\section{Quasistatic dynamics}
\label{quasistatic}

A key property of the AOUP model is that the self-propulsion force is entirely decoupled from the environment. As a result, one can integrate the equation of motion for $f$ independently of the position $x$:
\begin{align}
f(t) = f(0) + \int_0^t dt' \frac{\sqrt{D}}{\tau}\xi(t') e^{(t'-t)/\tau}
\label{eq:aoup_f}
\end{align}

\bigskip
The other key ingredient of our theory is an approximation: in the persistent limit ($\tau\rightarrow\infty$), the self-propulsion force $f$ varies quasistatically, \ie, we can solve equation~\eqref{eq:eom1} as if $f$ were constant and discard the transient part of the solution on the basis that its duration is negligible compared to the time scale over which $f$ varies. 

\begin{figure}
	\includegraphics[width=\linewidth]{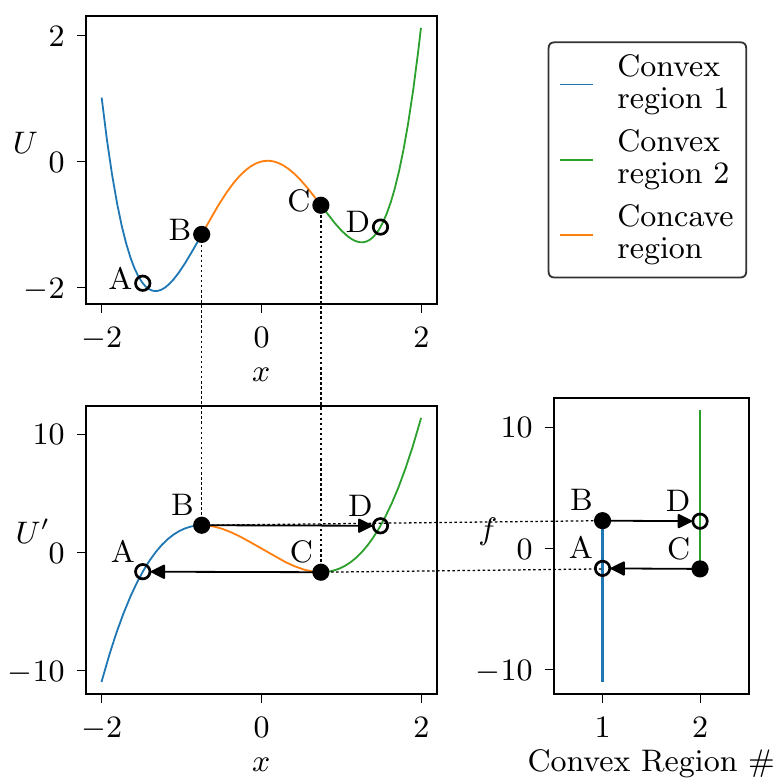}
	\caption{Quasistatic dynamics in a double well potential.}
	\label{fig:dynamics}
\end{figure}

\newcommand{\suchthat}{{\ |\ }}
Setting $f$ constant in equation~\eqref{eq:eom1} yields a first order ordinary differential equation for $x$ with no explicit time dependence. A particle initially at $x_0$ then moves in the direction indicated by the sign of $f-U'(x_0)$ until it reaches the next stable fixed point $x^*$, namely $x^*=\min\{x>x_0 \suchthat U'(x)=f \text{ and } U''(x)>0 \}$ if $f>U'(x_0)$ or $x^*=\max\{x<x_0 \suchthat U'(x)=f \text{ and } U''(x)>0 \}$ if $f<U'(x_0)$. In steady-state ($\dot{x}=0$) the particle simply remains at that fixed point. 
Graphically, the particle moves along the $U'(x)$ curve shown in the bottom left panel of figure~\ref{fig:dynamics} towards the horizontal line $U'=f$. If it is below the line, it moves to the right. If it is above the line, it moves to the left. If it reaches an intersection point, it stops. 

The quasistatic approximation assumes permanent steady-state, \ie, $U'(x)=f$ and $U''(x)>0$ are satisfied at all times. When the self-propulsion force changes, the particle jumps instantaneously from the fixed point corresponding to its old self-propulsion force to the one corresponding to its new self-propulsion force. If the potential is convex, the location is entirely determined by the current self-propulsion force through $x=(U')^{-1}(f)$. Conversely, if the potential is locally concave~\footnote{We do not consider potentials that are concave everywhere. They have no stable fixed point, causing particles to run away to $x=\pm\infty$.}, 
$U'$ is not monotonic and there are multiple locations with the same external force $-U'$, thus multiple possible fixed points for some values of $f$. Which of those fixed points the particle jumps to then depends on its previous location. 

The bottom left panel of figure~\ref{fig:dynamics} illustrates this situation in the case of a double well potential. $B$ and $C$ are the inflexion points of the potential. Each marks the boundary between the central concave region and one of the two convex regions. A particle with $U'_C<f<U'_B$ has two possible stable fixed points, one between $A$ and $B$ and one between $C$ and $D$. A particle previously left of $B$ jumps to the stable point between $A$ and $B$. A particle previously right of $C$ jumps to the stable point between $C$ and $D$. In other words, the particle stays in the convex region it was previously in. When reaching the end of a convex region, the particle jumps to a new location with the same $U'$ in a different convex region. A particle initially at $B$ whose self-propulsion force increases just past $U'_B$ jumps to $D$. A particle at $C$ whose self-propulsion force decrease just below $U'_C$ jumps to $A$. 

The bottom right panel of figure~\ref{fig:dynamics} shows an alternate representation of the problem that will prove most useful to derive the density profile. It is motivated by the observation that the dynamics of $f$ is much simpler to study than that of $x$. Most notably, it does not depend on the potential. In convex potentials, the one-to-one mapping $f=U'(x)$ means one can solve in $f$-space without worrying about $x$, then port the solution to $x$-space at the very end. A similar approach is possible in nonconvex potentials. However, for the mapping to remain one-to-one we must keep track of which convex region the particle is in as well as its self-propulsion force. In other words, we must solve for $f$ (or its probability distribution) in the space represented in the bottom right panel of figure~\ref{fig:dynamics}. Each vertical line corresponds to a convex region. The horizontal arrows correspond to the jumps the particle experiences when it reaches the end of a region. The concave region is not represented because it contains no stable fixed point. 

Finally, the dotted lines connecting the three panels of figure~\ref{fig:dynamics} highlight the way nonconvexity implies overlaps between the convex regions and jumps from the end of one region to the interior of another. 


\section{Density profile in a convex potential}
\label{convex}

To obtain the density profile $\rhox(x,t)$, we first solve for the density profile $\rhof(f,t)$ in the space of the self-propulsion force, \ie, the probability distribution of $f$, then use the one-to-one mapping $f=U'(x)$, whose bijectivity is guaranteed by the convexity of the potential, to get $\rhox(x,t)$. Applying standard It\=o calculus to equation~\eqref{eq:eom2} yields the usual Ornstein-Uhlenbeck Fokker-Planck equation:
\begin{align}
\partial_t \rhof = \partial_f \left( 
     \frac1\tau f \rhof + \frac{D}{\tau^2} \partial_f \rhof
     \right)
\label{eq:fp_f_convex}
\end{align}
The normalized steady-state solution is Gaussian:
\begin{align}
\rhof(f) = \sqrt{\dfrac{\tau}{2\pi D}}\ \exp\left[-\frac{\tau f^2}{2D}\right]
\label{eq:profile_f_convex}
\end{align}
A simple change of variable then yields the steady-state density profile:
\begin{align}
\rhox(x) = \frac{df}{dx}\, \rhof(f) = 
\dfrac{U''(x)\sqrt{\tau}}{\sqrt{2\pi D}}\  \exp\left[-\frac{\tau U'(x)^2}{2D}\right]
\label{eq:profile_convex}
\end{align}
In the rest of the paper we refer to those convex potential solutions as $\rhofc(f)$ and $\rhoxc(x)$.
\myskip

\emph{Comparison with UCNA. }
Before moving on to nonconvex potentials, it is instructive to compare this result with the equilibrium mapping prediction from the unified colored noise approximation (UCNA)~\cite{Maggi2015} \footnote{One may also use the Fox approximation, which yields the same steady-state profile~\cite{Wittmann2017}.}:
\newcommand{\thr}[1]{{\color{blue}#1}}
\newcommand{\qst}[1]{{\color{greenA}#1}}
\begin{align}
\rhox(x) = \dfrac{1}{Z}\, \Big[\thr{1}+\qst{\tau U''(x)}\Big]\,\exp\left[\thr{-\dfrac{U(x)}{D}} \qst{-\dfrac{\tau U'(x)^2}{2D}}\right]
\label{eq:profile_ucna}
\end{align}
where $Z$ is a normalization constant. This expression is exact in both the thermal limit ($\tau\rightarrow0$) and the persistent (or quasistatic) limit ($\tau\rightarrow\infty$). In the termal limit, it reduces to the usual Boltzmann weight: $\rhox(x)=\exp[-U(x)/kT]/Z$ where $k$ is the Boltzmann constant and $T=D/(k\mu)$ is the effective temperature. In the persistent limit, equation~\eqref{eq:profile_ucna} reduces to equation~\eqref{eq:profile_convex}, confirming the exactness of our quasistatic approach when $\tau\rightarrow\infty$. Interestingly, every term in~\eqref{eq:profile_convex} comes either from the thermal limit (blue terms) or the quasistatic limit (green terms). This decomposition makes sense if one recalls that the UCNA approximation is designed to be exact in both limits, and merely hopes that the intermediate $\tau$ regime can be inferred by extrapolating between them. What it does hightlight, though, is the need for a persistent limit result in nonconvex potentials, even if the gap between the thermal and persistent regimes cannot be bridged yet.


\section{Density profile in a nonconvex potential}
\label{nonconvex}

\subsection{Dynamics}

In a nonconvex potential, the mapping $f=U'(x)$ is not one-to-one anymore. We cannot simply derive $\rhox(x)$ from $\rhof(f)$ by change of variable. On the other hand, the mapping remains one-to-one within each convex region, and concave regions are empty because they host no stable fixed point. The solution, therefore, is to solve for the distribution $\rhof_i(f)$ in each convex region $i$. The dynamics of $\rhof_i(f)$ has two contributions: changes in the self-propulsion force of the particles within the convex region, and particles jumping from one region to another as illustrated in figure~\ref{fig:dynamics}. The first contribution alone leads to equation~\eqref{eq:eom2}. The second contribution takes the form of a point sink at each edge of each region (\ie, at every extremum of $U'$), where particles jump off, and a point source where the jump lands. 
Let $k$ index the jumps. Let $f_k$ be the value of $U'$ at extremum $k$. Let $J_k$ be the number of particles jumping from extremum $k$ per unit time. Let $\epsilon_{ik}=-1$ if jump $k$ starts in region $i$, $+1$ if jump $k$ lands in region $i$, and $0$ otherwise. Then:
\begin{align}
\partial_t \rhof_i = 
    \partial_f \left( \frac1\tau f \rhof_i + \frac{D}{\tau^2} \partial_f \rhof_i \right)
    + \sum_k \epsilon_{ij} J_k \delta(f-f_k)
\label{eq:fp_f_nonconvex}
\end{align}
Note that the probability density of $f$ regardless of the region, $\rhof(f)=\sum_i \rhof_i(f)$, still obeys equation~\eqref{eq:eom2} because each jump appears twice with opposite signs, once as a sink in the region where the jump originates, and once as a source in the region where it lands (same $f_k$, same $J_k$, opposite $\epsilon_{ik}$). In other words, $\rhof(f)$ is equal to the convex potential solution $\rhofc(f)$ from equation~\eqref{eq:profile_f_convex}.

Since every particle crossing the edge of a convex region jumps to a different convex region, the jump rate $J_k$ is equal to the probability flux through the edge from the interior of the region:  
\begin{align}
J_k = - \frac1\tau f \rhof_{i(k)}(f_k) - \frac{D}{\tau^2} \partial_f \rhof_{i(k)}(f_k)
\label{eq:J_p_relationship}
\end{align}
where $i(k)$ is the region where the jump originates. Substituting equation~\eqref{eq:J_p_relationship} into equation~\eqref{eq:fp_f_nonconvex} yields a closed set of coupled partial differential equations for the distributions $\rhof_i$. This is a convenient form to, \eg, integrate the dynamics numerically. 

For the purpose of deriving the steady-state density profile, however, it is more convenient to disregard equation~\eqref{eq:J_p_relationship} and treat the $J_k$'s as unknowns for a little longer.


\subsection{Steady-state}

To get the steady-state solution we set $\partial_t \rhof_i=0$ in equation~\eqref{eq:fp_f_nonconvex} and integrate once with respect to $f$:
\begin{align}
\frac1\tau f \rhof_i + \frac{D}{\tau^2} \partial_f p_i
= -\sum_k \epsilon_{ik} J_k \Theta(f-f_k) + a_i
\label{eq:fp_f_nonconvex_steady}
\end{align}
where $\Theta$ is the Heaviside function and $a_i$ is an integration constant. 
Next we rewrite the left-hand side as $\dfrac{D\rhofc}{\tau^2}\partial_f\left(\dfrac{\rhof_i}{\rhofc}\right)$ where $\rhofc(f)$ is the convex potential solution from equation~\eqref{eq:profile_f_convex}, multiply by $\tau^2/(D\rhofc)$, and integrate with respect to $f$ to get
\begin{align}
\rhof_i(f) = \left( b_i -\sum_{f_k<f} \epsilon_{ik} \dfrac{\tau^2 J_k}{D}\int_{f_k}^f \frac{du}{\rhofc(u)}\right) \rhofc(f)
\label{eq:f_nonconvex}
\end{align}
where $a_i$ has been set to $0$ to ensure $\lim_{f\rightarrow\infty} \rhof_i<\infty$ and $b_i$ is a new integration constant. Assuming the potential $U$ and the self-propulsion parameters $\tau$ and $D$ are known, the only unknowns left in equation~\eqref{eq:f_nonconvex} are the $b_i$'s and the $J_k$'s, which we must now solve for. Since equation~\eqref{eq:f_nonconvex} is linear in the unknowns, we need as many linearly independent equations as there are unknowns. 

A first set of equations comes from imposing either that $\rhof_i(f)$ can't diverge at $+\infty$ or that the total jump rate of any region has to be zero for the region's population to remain constant. Both yield $\sum_k \epsilon_{ik}J_k=0$. 

Another set of equations comes from the boundary conditions at each end of each region. The most common boundary condition is $\rhof_i=0$ at the boundary between a convex region and a convex region, which follows from imposing finite jump rates. However, a confining potential also has two convex regions that extend to $f=\pm\infty$, one on each side. Such a region has a range of $f$ in which it is the only available region. In that case the boundary condition is $\rhof_i(f)=\rhofc(f)$, which can be written at any $f$ in that range. Some types of potentials (\eg, nonconfining or noncontinuous) may require other types of boundary conditions, however we do not discuss them in this paper.

Those two sets of equations are typically sufficient to solve for the $b_i$'s and the $J_k$'s, thus for the self-propulsion force distribution in each convex region through equation~\eqref{eq:f_nonconvex}. 
The mapping to position-space $f=U'(x)$ is one-to-one in each convex region, therefore
$\rhox(x)=\frac{df}{dx}\rhof_i(f)=U''(x)\,\rhof_i[U'(x)]$ for $x$ in region $i$. 
Finally, the full density profile is 
\begin{align}
\rhox(x) = 
\begin{cases}
0 & \text{if } U''(x)<0 \\
U''(x)\rhof_{i(x)}[U'(x)] & \text{if } U''(x)>0
\end{cases}
\label{eq:nc_mapping}
\end{align}
where $i(x)$ is the index of the convex region $x$ belongs to (when $U''(x)>0$) and $\rhof_{i(x)}$ is the density in $f$-space in that region. 


\subsection{Double well potential}

In the case of a double well potential like the one shown in figure~\ref{fig:dynamics}, there are only two convex regions and two jumps. 
Region $1$ extends from $x=-\infty$ to $x=x_B$, or, in $f$-space, from $f=-\infty$ to $f=U'_B$. 
Region $2$ extends from $x=x_C$ to $x=\infty$ and from $f=U'_C$ to $f=\infty$. 
One jump goes from $B$ to $D$ with current $J_B$ and force $f=U'_B=U'_D$. 
The other goes from $C$ to $A$ with current $J_C$ and force $f=U'_C=U'_A$. 
Equation~\eqref{eq:f_nonconvex} reads
\begin{widetext}
\begin{align}
\frac{\rhof_1(f)}{\rhofc(f)} & = b_1
- J_C \Theta(f-U'_C) \int_{U'_C}^f \frac{\tau^2 du}{D\,\rhofc(u)}
+ J_B \Theta(f-U'_B) \int_{U'_B}^f \frac{\tau^2 du}{D\,\rhofc(u)}
\label{eq:doublewell1}
\\
\frac{\rhof_2(f)}{\rhofc(f)} & = b_2
+ J_C \Theta(f-U'_C) \int_{U'_C}^f \frac{\tau^2 du}{D\,\rhofc(u)}
- J_B \Theta(f-U'_B) \int_{U'_B}^f \frac{\tau^2 du}{D\,\rhofc(u)}
\label{eq:doublewell2}
\end{align}
\end{widetext}
The total jump rate equations are $-J_C+J_B=0$ (region $1$) and $J_C-J_B=0$ (region $2$). 

\begin{figure}
	\includegraphics[width=\linewidth]{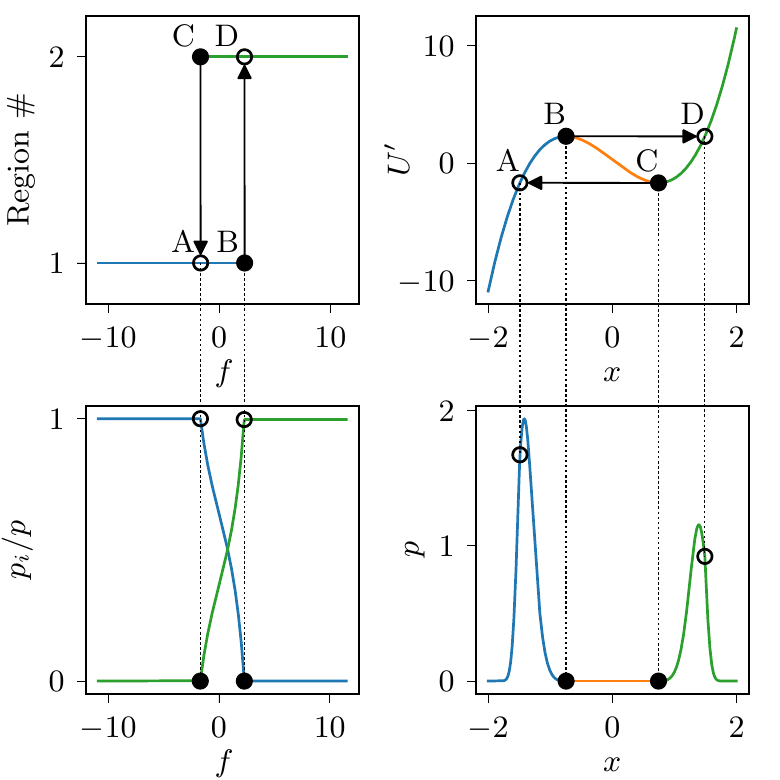}
	\caption{Density profile in the double well of figure~\ref{fig:dynamics}.
	The colors obey the legend of figure~\ref{fig:dynamics}.
	Top left: Self-propulsion force space showing the two convex branches, their overlap, and the jumps between them. Same as the bottom right panel of figure~\ref{fig:dynamics} except the axes have been swapped.
	Top right: Derivative of the potential showing the convex regions and the jumps between them. Same as the bottom left panel of figure~\ref{fig:dynamics}.
	Bottom left: Steady-state density profile $\rhof(f)$ in each convex branch in self-propulsion force space.
	Bottom right: Steady-state density profile $\rhox(x)$ in position space (\ie, the actual density profile).
	}
	\label{fig:density}
\end{figure}

If $f<U'_A$, the only available stable fixed point is in region~$1$. As a result, every particle with $f$ in that range must be in region~$1$: $\forall f<U'_A,\ \rhof_1(f)=\rhofc(f)$. In particular, $\rhof_1(U'_A)=\rhofc(U'_A)$, which we use as the ``left end'' boundary condition for region~$1$. 
Conversely, a particle with $f<U'_A$ cannot be in region~$2$: $\forall f<U'_A,\ \rhof_2(f)=0$. In particular $\rhof_2(U'_C)=0$, which is our left end boundary condition for region~$2$. 
Similarly, $\forall f>U'_D,\ \rhof_1(f)=0$ and $\rhof_2(f)=\rhofc(f)$, which yields the right end boundary conditions $\rhof_1(U'_B)=0$ and $\rhof_2(U'_D)=\rhofc(U'_D)$.

After dividing the boundary condition equations by $\rhof(f)$, the complete system (jump rates and boundary conditions) reads 
\begin{align*}
-J_C + J_B & = 0 \\
 J_C - J_B & = 0 \\
b_1 & = 1 \\
b_1 - J_C \int_{U'_C}^{U'_B} \frac{\tau^2 du}{D \rhofc(u)} & = 1 \\
b_2 & = 0 \\
b_2 + J_C \int_{U'_C}^{U'_B} \frac{\tau^2 du}{D \rhofc(u)} & = 1
\end{align*}
There are four unknowns and six equations but only four linearly independent equations. The (unique) solution is 
$J_B=J_C=\left(\int_{U'_C}^{U'_B} \frac{\tau^2 du}{D \rhofc(u)}\right)^{-1}$, 
$b_1=1$, $b_2=0$. The corresponding density in $f$-space is
\begin{align}
\frac{\rhof_1(f)}{\rhofc(f)} & = 
\begin{cases}
1 & \text{ if } f<U'_A \\
\dfrac{\int_{U'_B}^{f} du\, e^{\frac{\tau u^2}{2D}}}{\int_{U'_B}^{U'_A} du\, e^{\frac{\tau u^2}{2D}}}
& \text{ if } U'_A<f<U'_B \\
0 & \text{ if } f>U'_B
\end{cases}
\\
\frac{\rhof_2(f)}{\rhofc(f)} & = 
\begin{cases}
0 & \text{ if } f<U'_C \\
\dfrac{\int_{U'_C}^{f} du\, e^{\frac{\tau u^2}{2D}}}{\int_{U'_C}^{U'_D} du\, e^{\frac{\tau u^2}{2D}}}
& \text{ if } U'_C<f<U'_D \\
1 & \text{ if } f>U'_D
\end{cases}
\end{align}
where $\rhofc$ is given by equation~\eqref{eq:profile_f_convex}. The density profile in position space follows by applying the change of variable $f=U'(x)$ in each region separately: 
\begin{align}
\frac{\rhox(x)}{\rhoxc(x)} & = 
\begin{cases}
1 & \text{ if } x<x_A \text{ or } x>x_D \\
\dfrac{\int_{U'_B}^{U'(x)} du\, e^{\frac{\tau u^2}{2D}}}{\int_{U'_B}^{U'_A} du\, e^{\frac{\tau u^2}{2D}}}
& \text{ if } x_A<x<x_B \\
0 & \text{ if } x_B<x<x_C \\
\dfrac{\int_{U'_C}^{U'(x)} du\, e^{\frac{\tau u^2}{2D}}}{\int_{U'_C}^{U'_D} du\, e^{\frac{\tau u^2}{2D}}}
& \text{ if } x_C<x<x_D 
\end{cases}
\label{eq:doublewell_prediction}
\end{align}
where $\rhoxc$ is given by equation~\eqref{eq:profile_convex}.


\subsection{Arbitrary confining potential}
\label{arbitrary_confining_potential}

Consider a potential whose derivative is continuous and goes from $-\infty$ to $+\infty$. The potential is confining in the sense that, for any probability $P$, no matter how close to $1$, there is a finite region of space that contains the particles with probability $P$. More importantly for the density profile, there is a stable fixed point at any self-propulsion force $f$.

The derivation of the steady-state density profile follows the same steps as in the double well case above. 
There is now an arbitrary number $N$ of convex regions. The $(i,f)$ representation in the top left panel of figure~\ref{fig:density} has $N$ branches. Let region $1$ be the (only) one that extends to $f=-\infty$ and region $N$ the (only) one that extends to $f=+\infty$. The left boundary condition for region $1$ yields $b_1=1$ the same way it did for the double well's region $1$. The left boundary condition for every other region yields $a_i=0$ the same way it did for the double well's region $2$. Similarly, the right boundary conditions are $p_i=0$ for $i<N$ and $p_N=p$. They are all evaluated at the region's rightmost jump $f_i^R\equiv\max_k |\epsilon_{ik}| f_k$, which coincides with either the end of the region ($i<N$) or the point where region $N$ no longer overlaps with any other region. After including the total jump rate equations we get:
\begin{align}
a_i & = \delta_{1,i} \label{eq:jsys1} \\
\sum_{k} \epsilon_{ik} J_k & = 0 \label{eq:jsys2} \\
\delta_{1,i} + \sum_{k} \epsilon_{ik} J_k \int_{f_k}^{f_i^R} \frac{\tau du}{D\,\rhofc(u)} & = \delta_{N,i} \label{eq:jsys3}
\end{align}
$\forall i\in[1,N]$ where $\delta$ is the Kronecker delta.

Equation \eqref{eq:jsys1} is already a solution for the $a_i$'s. Equations \eqref{eq:jsys2} and \eqref{eq:jsys3} each provide $N$ linear equations for the $J_k$'s. Since $\sum_i \epsilon_{ik}J_k=0$ (the source and sink corresponding to the same jump cancel), summing equation \eqref{eq:jsys2} over $i$ yields zero and \eqref{eq:jsys2} provides at most $N-1$ linearly independent equations. Similarly, summing equation \eqref{eq:jsys3} over $i$ yields 
\begin{multline}
\sum_{i,k} \int_{f_k}^{f_i^R} \frac{\tau du}{D \rhofc(u)} \sum_k \epsilon_{ik} J_k = \\
\sum_i \int_0^{f_i^R} \frac{\tau du}{D \rhofc(u)} \sum_k \epsilon_{ik} J_k \\
- \sum_k \int_0^{f_k} \frac{\tau du}{D \rhofc(u)} \sum_i \epsilon_{ik} J_k
\label{eq:linear_constraint}
\end{multline}
The first inner sum on the right-hand side is \eqref{eq:jsys2}. The second inner sum is zero because $\sum_k \epsilon_{ik}J_k=0$. Thus equations~\eqref{eq:jsys2}-\eqref{eq:jsys3} provide at most $2N-2$ linearly independent equations. This is also the number of jumps, thus of jump rates: one from each of regions $1$ and $N$, and two from each of the other regions. Numerically constructing the system \eqref{eq:jsys2}-\eqref{eq:jsys3} for some specific potentials suggests that eliminating one equation from \eqref{eq:jsys2} and one equation from \eqref{eq:jsys3} does typically yield a nonsingular system that can be inverted to obtain the jump rates, then the density profile by equation~\eqref{eq:nc_mapping}. 
In other words, this is a systematic route to deriving the density profile in an arbitrary external potential.


\section{Discussion}
\label{discussion}

Regardless of the specifics of the confining potential, the structure of the density profile derived in section~\ref{arbitrary_confining_potential} contains valuable information. Let us first summarize the result: 
\begin{multline}
\rhox(x) = \rhoxc(x) \times \Theta[U''(x)] \\
\times \left( \delta_{i(x),1} -\sum_{f_k<f} \epsilon_{i(x),k} \dfrac{\tau^2 J_k}{D}\int_{f_k}^f \frac{du}{\rhofc(u)}\right).
\label{eq:summary1}
\end{multline}
$U(x)$ is the external potential. The $f_k$'s are the values of the external force at the inflexion points of $U(x)$. $\rhofc$ and $\rhoxc$ are given by equations~\eqref{eq:profile_f_convex}-\eqref{eq:profile_convex}. $i(x)$ is the label of the convex region $x$ is located in. $\epsilon_{ik}$ is $1$ if region $i$ ends at $k$, $-1$ if another convex region spills into region $i$ at $f_k$, and $0$ otherwise. The jump rates $J_k$ obey
\begin{align}
\sum_{k} \epsilon_{ik} J_k & = 0 \label{eq:summary2} \\
\delta_{1,i} + \sum_{k} \epsilon_{ik} J_k \int_{f_k}^{\max_k |\epsilon_{ik}| f_k} \frac{\tau du}{D\,\rhofc(u)} & = \delta_{N,i} 
\label{eq:summary3}
\end{align}
\myskip

\emph{Features inherited from convex potentials.} 
The first term on the right-hand side of \eqref{eq:summary1}, $\rhoxc(x)$, is the density profile one would observe in a potential that would be convex everywhere and match $U$ around the point of interest $x$. 
From this term the density profile inherits most of the properties of convex-potential density profiles. 
Perhaps the most striking one is that the density does not depend on $U$ at all, only on $U'$ and $U''$.
In particular, particles do not accumulate where $U$ is lowest but where $\tau U'^2-D\log U"^2$ is lowest, \eg, at walls~\cite{Maggi2015,Fily2017}. 
\myskip

\emph{Emptiness of concave regions.} 
The second term on the right-hand side of \eqref{eq:summary1}, $\Theta[U''(x)]$, means that concave regions of $U$ are empty. This is because the particles only get stuck in convex regions. The time for which they remain stuck is controlled by their persistence time $\tau$, whereas the time they spend passing through concave regions is controlled by their self-propulsion speed $v_0$ and the size of the region. In the persistent limit $\tau\rightarrow\infty$, particles spend a vanishingly small fraction of their time in concave regions which end up empty.
\myskip

\emph{Importance of the potential's inflection points.}
The third and last term on the right-hand side of \eqref{eq:summary1} captures the complex flow of particles between the convex regions of the potential. Strikingly, this term only depends on the potential $U$ through 1) the value of its derivative $U'$ at its inflection points (the $f_k$'s), 2) which convex region spills into which convex region (the $\epsilon_{ik}$'s), and 3) the rate at which particles jump between regions (the $J_k$'s). The $J_k$'s in turn only depend on the $f_k$'s and the $\epsilon_{ik}$'s through equations~\eqref{eq:summary2}-\eqref{eq:summary3}, and the $\epsilon_{ik}$ are largely determined by the pair of $f_k$'s associated to each convex region. In other words, at the end of the daya it's all about the inflection points, more specifically the value of the external force at each inflection point.
\myskip

\emph{Nonlocality of the nonconvex solution. }
In a convex potential, the dependence of the density on the external potential is local: the density at position $x$ only depends on the potential through its value and the values of its first two derivatives at $x$ (see equation~\eqref{eq:profile_convex}). In a nonconvex potential, the density profile acquires a nonlocal dependence on the potential through the $f_k$'s, $J_k$'s, and $\epsilon_{ik}$'s. As we pointed out above, this dependence ends up being mostly on the $f_k$'s, \ie, the value of the external force at the inflection points of the potential, however far they may be from the point of interest.


\section{Conclusion}

Despite their considerable successes, the inability of the equilibrium mapping techniques initiated by \cite{Farage2015,Maggi2015} to provide insight into the density profile of very persistent self-propelled particles under nonconvex confinement is a major issue as this is precisely where uniquely active phenomena such as rectification occur. 
This paper describes an alternative approach which, although much less general, tackles the very regime in which previous methods fail most dramatically. Specifically, we derived the density profile of an active Ornstein-Uhlenbeck active particle in an arbitrary confining potential in one dimension in the persistent limit. We found it to exhibit very different properties from previous equilibrium mapping predictions, including a vanishing density in regions where the potential is concave, a nonlocal dependence of the density on the external potential, and the disproportionate influence of the external force at the potential's inflection points.
If this approach can be extended to finite persistence times and higher dimensions, it will no doubt offer significant novel theoretical insight into the surprising physics of confined active matter.


\bibliography{1d_gcn}

\end{document}